\documentclass[11pt]{article}

\usepackage{epsfig}
\usepackage{amssymb}
\usepackage{fullpage}
\usepackage{graphicx}

\title{Imaging with two-axis micromirrors}

\bigskip

\author{R. Andrew Hicks\\
Department of Mathematics, Drexel University \\ 3141 Chestnut St., Philadelphia, PA 19104, USA\\ \ \\
Vasileios T. Nasis\\
Department of Electrical and Computer Engineering, Drexel University \\ 3141 Chestnut St., Philadelphia, PA 19104, USA\\ \ \\
Timothy P. Kurzweg\\Department of Electrical and Computer Engineering, Drexel University \\ 3141 Chestnut St., Philadelphia, PA 19104, USA
}
\date{}

\begin{document}

\maketitle

\begin{abstract}
We demonstrate a means of creating a digital image by using a two axis tilt micromirror to scan a scene. For each different orientation we extract a single grayscale value from the mirror and combine them to form a single composite image. This allows one to choose the distribution of the samples, and so in principle a variable resolution image could be created. We demonstrate this ability to control resolution by constructing a voltage table that compensates for the non-linear response of the mirrors to the applied voltage.

\end{abstract} 


\noindent
Micro-Opto-Electro-Mechanical Systems (MOEMS) is a relatively new field which appears to have a multitude of applications. Perhaps the best known MOEMS is the Texas Instruments' Digital Micromirror Device (DMD), which is an NxN array of SRAM cells, each covered by a tilting mirror \cite{hornbeck98}. Each of these mirrors is either in a binary
``on" or ``off" state. The primary application of this device is for the projection of images. Optical switching is another application area for MOEMS.  Here, mirrors can eliminate the costly conversion from the optical domain to the electrical domain for switching. An example is Lucent's WaveStar LambdaRouter, which used an 8x8 array of 2-axis 600 $\mu m$ diameter micromirrors to achieve fiber array switching for 256 channels \cite{lucent01}. A later version of the array consisted of 16x16 mirrors. Each mirror may be individually actuated and can achieve 100,000 distinct states \cite{kim03photonics}. 

In this letter, we present our initial results on photographic imaging with micromirror arrays. Using a prototype  which employs a 16x16 Lucent micromirror array we have been able to obtain low resolution images with a single mirror, and illustrate a means of calibrating the mirrors for imaging purposes. Note that while the term ``imaging" is commonly associated with MOEMS, it is usually not used in the sense of photographic imaging \cite{moemsimagingiv}. Exceptions include work done by Nayar et al. with the DMD to extend dynamic range and work by Last et al. on  micro-cameras which use a 1-axis mirror to increase the field of view\cite{nayar04cvpr, last03}. Previous work by the authors described simulations and some manual experiments\cite{nasis06photonics}.

In conventional macroscopic photography there is the notion of image mosaicing, in which one combines two or more images with some common overlap to create a single image of higher resolution than its constituents. An omnidirectional example, which is the main motivation for this work, is that done by Kropp et al. as part of the MIT City Scanning Project\cite{kropp00cvpr}. In mosaicing, generally a camera is moved to obtain images. A conceivable alternative is to point the camera at a movable mirror\cite{nakeo01icip}. Here we propose the use of micromirrors for this purpose in conjunction with a video camera for  the purpose of obtaining one or more pixels from each image of the recorded video and combining them to form an image. One  characteristic of such a sensor is its ability to sample in a prescribed fashion. For example, if an image was recorded, then it should be possible to choose a region of the scene for closer inspection, and extract more samples from that region. This becomes particularly attractive for wide angle or panoramic imaging in which the total solid angle being imaged is large, and hence a relatively small number of pixels are allocated to a typical steradian.  For example, conventional panoramic imaging systems consisting of a curved mirror and video camera have the drawback that the resolution is generally non-uniform. While it is possible to design catadioptric systems that are equiresolution, these systems still lack the ability to ``zoom in" on an object of interest\cite{hicks05ao-equi}. If a curved mirror is imaged with a micromirror though, more pixels could be taken from the regions of interest. This could be especially useful for tracking and surveillance applications.
 
\begin{figure*}[t]
\centerline{\includegraphics[width=10cm]{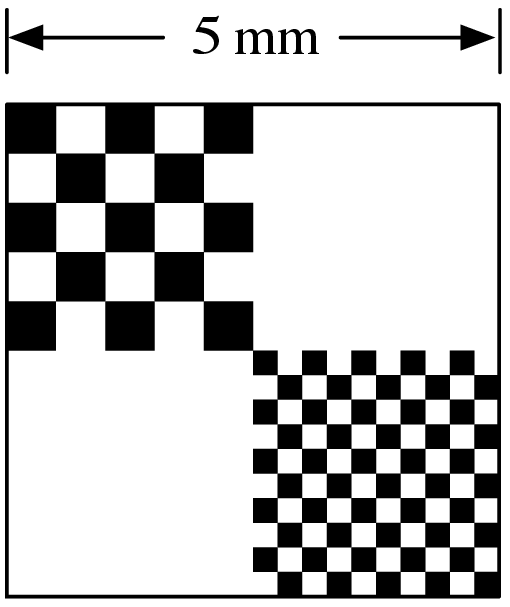}}
\caption{The fundamental unit of our test pattern.  }
\label{fig:unit}
\end{figure*}

\begin{figure*}[t]
\centerline{\includegraphics[width=10cm]{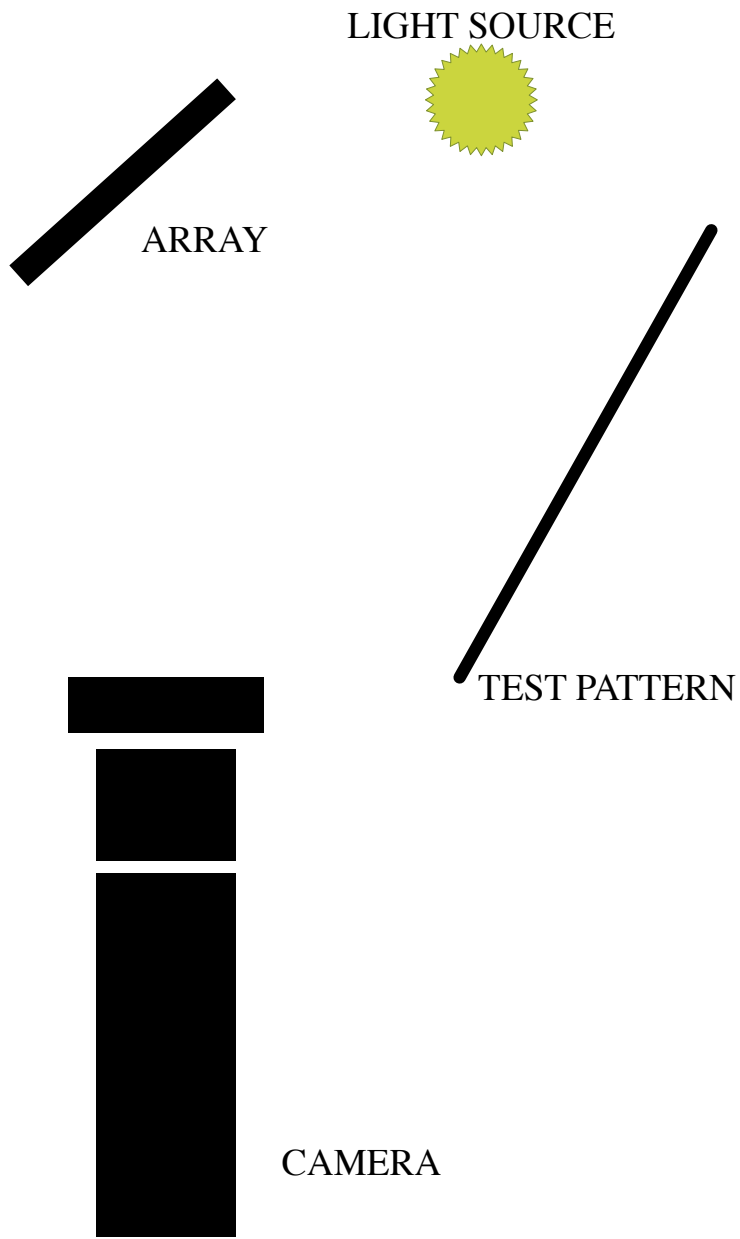}}
\caption{The relationship between the camera, the array, the illumination source and the test pattern.}
\label{fig:setup}
\end{figure*}

\begin{figure*}[t]
\centerline{\includegraphics[width=10cm]{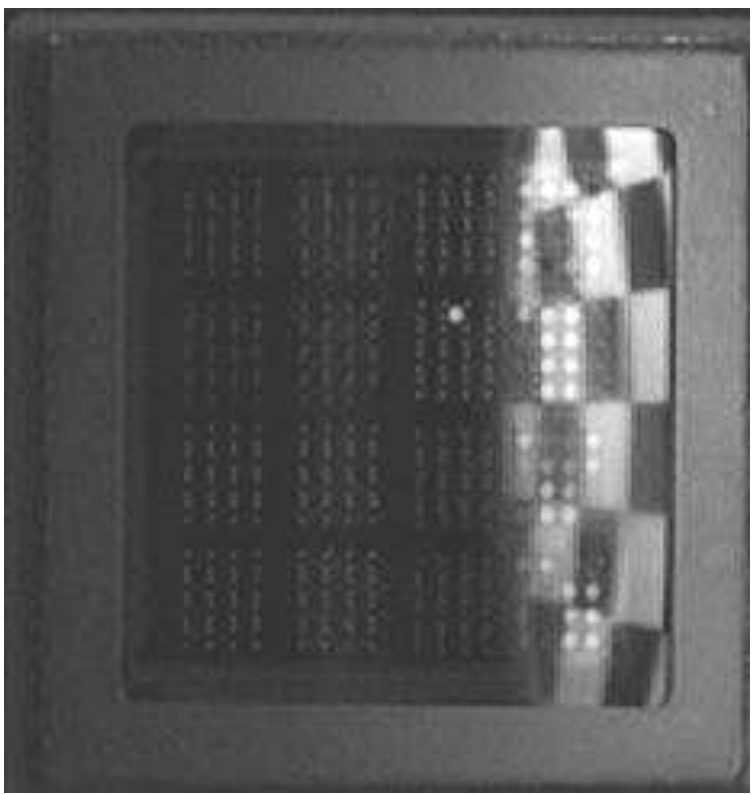}}
\caption{The Lucent 16x16 array. The checkerboard pattern to the right  is the reflection of a test pattern off of the sapphire cover. To the left of this we see a single white spot, which is a mirror, tilted to reflect a white portion of the test pattern. It was this gray value that was collected from each frame.}
\label{fig:array}
\end{figure*}

\begin{figure*}[t]
\centerline{\includegraphics[width=10cm]{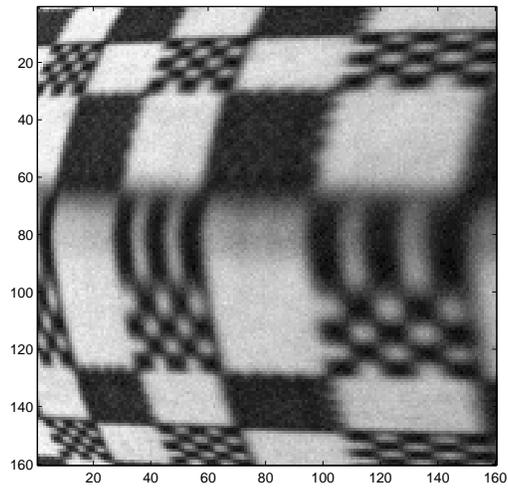}}
\caption{ A 160x160 image created by scanning the test pattern with a single mirror and uniformly spaced voltages.}
\label{fig:data}
\end{figure*}

\begin{figure*}[t]
\centerline{\includegraphics[width=10cm]{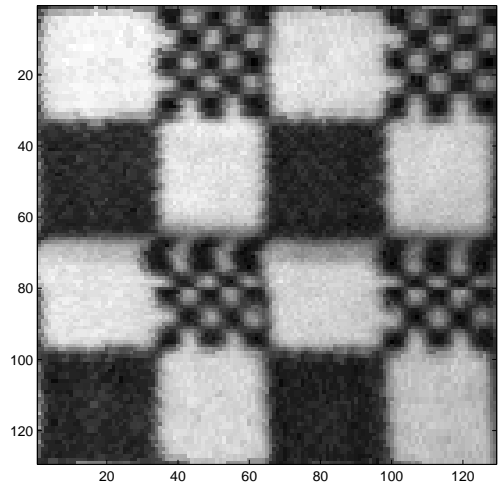}}
\caption{A 129x129 image obtained by using a nonlinear voltage table.}
\label{fig:uniform}
\end{figure*}

Through a loan from Lucent Technologies, the authors obtained a 16x16 mirror array and the associated control electronics. We have constructed images by scanning a test pattern consisting of 5 mm squares, each broken into four sub-squares, two of which contained smaller checkerboard patterns, with checkers .5 mm in the northwest sub-square and .25 mm in the southeast sub-square, as depicted Fig. \ref{fig:unit}. Using a single mirror we extracted a single pixel from a fixed single mirror of the array. Fig. \ref{fig:setup} is a schematic of our device and Fig. \ref{fig:array} is an image of the array. The Lucent mirrors have a $\pm 7$ degree tilt, with voltages varying from $\pm 120$. Scanning the test pattern with uniformly spaced voltages from -105 to -65 volts on one actuator and -79 to 80 volts on another for a fixed single mirror, we constructed a 160x160 composite grayscale image, depicted in Fig. \ref{fig:data}, by extracting a single pixel from each image. The non-linear response of the mirrors to the applied voltages results in a distorted image. Nevertheless, using this composite image, it was then  possible to build a non-linear table of voltages, which could sample the image uniformly. Thus we may control how the scene is sampled. The table was built using the fact that the voltages corresponding to the corners of the checkers were known. 81 sample points were chosen resulting in two 9x9 voltage tables. These were then expanded to 129x129 tables via bilinear interpolation and this table was then used to image the same test pattern. The result appears in Fig. \ref{fig:uniform}. In both images the .25 mm checkers are not visible, but the .5 mm checkers are, so clearly we are in some sense at the limit of the resolution of the device. These images were created using a conventional 640x480 video camera (Sony DFW-V300) and a 75 mm double gauss macro lens. Our illumination source was a 35W halogen lamp placed 7 cm from the test pattern, which was 6 cm away from the array. The F/\# of the lens could vary from F/4 to F/30 and we our data was collected at approximately F/25. The camera was 125 cm from the array, which was tilted at a 45 degree angle with the optical axis of the camera. See Fig. \ref{fig:setup} for a schematic representation.

A technical issue that had to be overcome was reflection from the glass cover. The Lucent array is covered with a sapphire window with an anti-reflection coating, which peaks at 1.4 $\mu$n. Thus in the visible wavelength we experienced considerable reflection from the glass cover, but by careful choice of orientation of the array and the test pattern we were able to minimize the reflected image from the single mirror that we used. 

A second technical problem arose due to the dead space between the mirrors. The camera/lens should be focused on the test pattern of course, but generally this meant that objects at the depth of the array were out of focus. As a result, the blur of the dead space made it impossible to resolve the mirrors unless the mirrors and the test pattern both were within the depth of field of the camera.

What are the advantages of imaging with micromirror arrays? Our work here is merely a proof of concept, and we reached the highest resolution using off the shelf optics immediately. Our ultimate vision though is that one could have a sensor in which data was being gathered simultaneously from all of the mirrors of the array at a high frame rate. Possibly several pixels could be extracted from each mirror. The array we used was certainly not optimized for this application and there were several technical issues that were not specific to the concept. 

In a very optimistic scenario, if one has a 640x480 array which was used for a 1 second exposure and a single pixel was extracted from each mirror in each frame at 30 frames per second, then the resulting image would have 9 megapixels. This assumes that each mirror could be made to image distinct regions of the scene, which is a technical issue {\it not} specific to our prototype. In other words, two adjacent mirrors will image practically the same solid angle. In order to avoid this problem supplementary optics could to be introduced. There is also the alternative possibility of ``interweaving" the pixel values, but this would seem to require an extremely precise calibration of the mirrors. On the other (even more optimistic) hand, if one could obtain 4 pixels from each mirror at 60 frames per second, the resulting image would have 72 megapixels. Thus an enormous omnidirectional image of a scene could be created, with a prescribed sampling density of the scene. In any case, the point is that the performance limits of such a device would vary greatly with a number of parameters, but in some scenarios the results are quite interesting. Nevertheless, an advantage of imaging with micromirrors is that as frame rates and camera resolutions increase, the same principles apply. The possible applications of this concept will depend heavily upon technological advances in several areas.

\section*{Acknowledgments}
The authors would like to thank Vladimir Aksyuk, John Gates and and Rick Papazian of Lucent Technologies/Bell Laboratories, G.K. Ananthasuresh of the Indian Institute of Science, Bangalore, Jungsang Kim of Duke University ECE, and Seth Teller of MIT EECS. This work was supported by NSF-DMS-0211283 and NSF-IIS-0413012.


\bibliographystyle{unsrt}

\end{document}